\documentclass[prl,aps,showpacs,preprint]{revtex4}

\usepackage{epsfig}

\begin{document}

\title{Drastic Emergence of Huge Negative Spin-Transfer Torque in Atomically Thin Co Layers}
\author{Soong-Geun Je,$^{1,\dagger}$ Sang-Cheol Yoo,$^{1,2}$ Joo-Sung Kim,$^{1}$ Yong-Keun Park,$^{1,2}$ Min-Ho Park,$^{1}$ Joon Moon,$^{1}$ Byoung-Chul Min,$^{2}$ and Sug-Bong Choe$^{1,*}$}
\affiliation{$^1$Department of Physics and Institute of Applied Physics, Seoul National University, Seoul 151-742, Republic of Korea}
\affiliation{$^2$Center for Spintronics Research, Korea Institute of Science and Technology, Seoul 136-791, Republic of Korea}
\affiliation{$^{\dagger}$Present Address: SPINTEC, CEA/CNRS/Univ. Grenoble Alpes, 38054 Grenoble, France and Institut Jean Lamour, UMR 7198 CNRS-Universit{\'e} de Lorraine, 54506 Vandoeuvre l{\`e}s Nancy, France}
\date{\today}

\begin{abstract}
Current-induced domain wall motion has drawn great attention in the last decades as the key operational principle of emerging magnetic memory devices.
As the major driving force of the motion, the spin-orbit torque on chiral domain walls has been proposed and extensively studied nowadays.
However, we demonstrate here that there exists another driving force, which is larger than the spin-orbit torque in atomically thin Co films.
Moreover, the direction of the present force is found to be opposite to the prediction of the standard spin-transfer torque, resulting in the domain wall motion along the current direction.
The symmetry of the force and its peculiar dependence on the domain wall structure suggest that the present force is, most likely, attributed to considerable enhancement of a negative nonadiabatic spin-transfer torque in ultra-narrow domain walls.
Careful measurements on the giant magnetoresistance manifest a negative spin polarization in the atomically thin Co films, which might be responsible for the negative spin-transfer torque.
\end{abstract}


\maketitle
Recent achievements in the current-driven domain wall motion (CIDWM) promise abundant opportunities in spintronic memory and logic devices~\cite{ParkinScience}.
It is now possible to attain high domain wall (DW) speeds up to nearly 1,000 m/s~\cite{YangNatNano}, low operation current densities down to 10$^9$ A/m$^2$~\cite{LeePRL}, and DW position accuracy better than 1 ${\mu}$m~\cite{ThomasScience}.

Despite these technological advances, the underlying mechanisms of the CIDWM remain to be clarified.
The most prominent example of the controversy is the DW motion along the current direction~\cite{LeePRL, KimAPEX, MironNatMater, LavrijsenAPL, MooreAPL}, which is opposite to the prediction of the original spin-transfer torque (STT) theory~\cite{BergerJAP, SlonczewskiJMMM}.
To explain such opposite motion, it has been proposed and demonstrated that the spin-orbit torque (SOT)---generated by the spin-Hall effect~\cite{LiuPRL} and/or the Rashba effect~\cite{MironNature}---can exert a force in the direction of the current on chiral DWs, of which the chirality is controlled by an in-plane magnetic field~\cite{HaazenNatMater}.
Recent observation of a built-in DW chirality~\cite{ChenPRL, JePRB} due to the Dzyaloshinskii-Moriya interaction (DMI) enables one to accomplish such DW motion~\cite{EmoriNatMater, RyuNatNano, FrankenSciRep, EmoriPRB, TorrejonNatComm, ThiavilleEPL}.
Therefore, the SOT combined with the DMI has been extensively studied as the underlying mechanism of the CIDWM.
On the other hand, the STT has been observed to be negligible in thin ($\sim$0.6 nm) magnetic layers~\cite{FrankenSciRep, EmoriPRB, CormierPRB}, possibly due to the very small current through the layers.

However, here we demonstrate that, in the films with ultra-thin ($\sim$0.3 nm) magnetic layers, the STT is significantly enhanced so that it is even larger than the SOT and thus, plays a significant role in the CIDWM.
A careful experimental analysis reveals that such significant enhancement can be caused by the formation of the ultra-narrow DWs of a few nanometers, with which the nonadiabatic STT is radically increased as suggested by the nonadiabatic STT theory~\cite{TataraPRL}.
In addition, the sign of the nonadiabatic component is found to be negative, opposite to the original STT theory, offering a new degree of freedom in the STT.

For this study, a series of 5-${\mu}$m-wide ferromagnetic strips made of Pt/Co/Pt films with various thicknesses of the Co and Pt layers was prepared.
In these films, the sign and magnitude of the SOT can be tuned by adjusting the thicknesses of the Pt layers~\cite{HaazenNatMater} and the DW chirality can be easily controlled by a small in-plane field due to the weak DMI~\cite{JePRB}.

\begin{figure}
\includegraphics[width=8.6 cm]{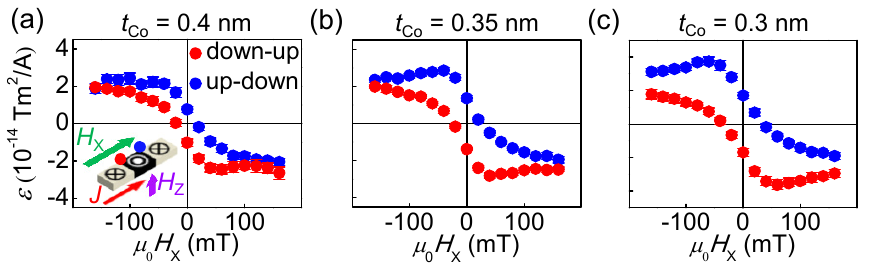}
\caption{
(a)-(c) $\varepsilon$ of the down-up (red, $\varepsilon^+$) and up-down (blue, $\varepsilon^-$) DWs with respect to $\mu_0 H_x$ for strips with $t_{\rm Co}$ of 0.4 (a), 0.35 (b) and 0.3 nm (c). The inset of (a) shows the illustration of the current-assisted field-driven DW motion for $\varepsilon$ measurement.
}
\label{fig1}
\end{figure}

The spin torque efficiency ${\varepsilon}$ of the CIDWM in these strips was then investigated.
Here, ${\varepsilon}$ is defined as a proportionality constant between the current density $J$ and the induced effective magnetic field $H_z^*$, based on the relation $H_z^* (J)={\varepsilon}J$~\cite{LeePRL}.
$H_z^*$ was measured from the DW creeping motion for two types of DW (i.e., down-up and up-down DWs) as depicted in the inset of Fig.~\ref{fig1}(a).
Figures~\ref{fig1}(a)-(c) show the plots of $\varepsilon$ with respect to the in-plane magnetic field $H_x$ for (2.5-nm Pt/$t_{\rm Co}$ Co/1.5-nm Pt) strips with different Co layer thickness $t_{\rm Co}$.
The epitaxial growth of the layers and the formation of a uniform magnetic layer without discontinuities on a length scale of DW were confirmed by a scanning transmission electron microscope and a magneto-optical Kerr effect microscope~\cite{comment2}.
The nominal film thicknesses were determined from a deposition rate ($\sim0.25$ {\AA}/sec). 
In each plot, the $\varepsilon^+$ (red) denotes $\varepsilon$ of the down-up DW and the $\varepsilon^-$ (blue) denotes that of the up-down DW.

The strip with $t_{\rm Co}=0.4$ nm, as shown in Fig.~\ref{fig1}(a), exhibits a very typical shape: Both $\varepsilon^+$ and $\varepsilon^-$ are saturated at high $|H_x|$ regimes through a gradual transition at small $|H_x|$ regime.
This typical shape is caused by the SOT because the SOT-induced efficiency $\varepsilon_{\rm SOT}$ is proportional to the $x$ component of the DW magnetization~\cite{HaazenNatMater}.
In addition, the horizontal shifts of $\varepsilon^+$ and $\varepsilon^-$ in opposite directions indicate the existence of a finite DMI, which generates an effective field $H_{\rm DMI}$ along the $x$ axis.
Therefore, the measurement of $\varepsilon^{\pm}(H_x)$ enables one to quantify both $H_{\rm DMI}$ and $\varepsilon_{\rm SOT}$, as demonstrated in the recent reports~\cite{FrankenSciRep, EmoriPRB}.

However, as $t_{\rm Co}$ is further reduced down to 0.35 and 0.3 nm, as shown by Figs.~\ref{fig1}(b) and (c), $\varepsilon^{\pm}(H_x)$ becomes largely deviated from the typical shape of $\varepsilon_{\rm SOT}$.
In these strips, $\varepsilon(H_x)$ does not show a monotonic saturation at large $|H_x|$ regimes, but shows a maximum at an intermediate $H_x$ regime.
This observation signals the existence of another contribution, distinct from the typical $\varepsilon_{\rm SOT}$.

\begin{figure}

\includegraphics[width=8.6 cm]{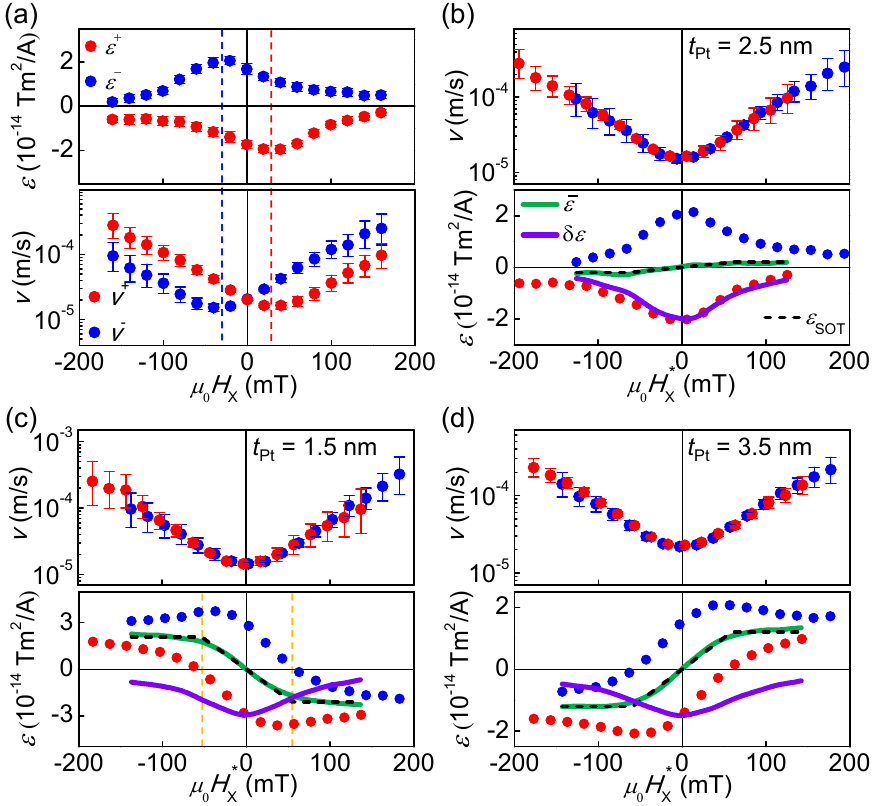}
\caption{
(a) $\varepsilon$ (upper panel) and $v$ (lower panel) plotted with respect to $\mu_0 H_x$ for the strip with $t_{\rm Pt}= 2.5$ nm. (b)-(d) $v$ (upper panels) and $\varepsilon$ (lower panels) plotted with respect to $\mu_0 H_x^*$ for strips of $t_{\rm Pt}=2.5$ (b), 1.5 (c) and 3.5 nm (d). The red (blue) symbol corresponds to the down-up (up-down) DW. For $v$, the applied field $\mu_0 H_z$ was $\pm1$ (b), $\pm3.5$ (c), and $\pm1$ mT (d).
}
\label{fig2}
\end{figure}

To see the origin of the deviation, another series of (2.5-nm Pt/0.3-nm Co/$t_{\rm Pt}$ Pt) strips with different Pt layer thickness $t_{\rm Pt}$ was examined.
The upper panel of Fig.~\ref{fig2}(a) shows $\varepsilon^{\pm}(H_x)$ of the symmetric strip (i.e., $t_{\rm Pt}= 2.5$ nm).
In this strip, the SOT is almost compensated due to the symmetric Pt layer structure~\cite{HaazenNatMater, comment1, comment2}.
However, the deviation is still observed, strongly suggesting again that the deviation does not come from the SOT and thus, there exists another origin that generates a large contribution in $\varepsilon$.

Because $\varepsilon$ is sensitive to the DW configuration~\cite{HaazenNatMater, EmoriNatMater, RyuNatNano, FrankenSciRep, EmoriPRB, TorrejonNatComm}, the DW configuration is independently identified by measuring the asymmetric field-driven DW speed $v$ with respect to ${H_x}$~\cite{JePRB}.
The lower panel of Fig.~\ref{fig2}(a) shows $v^+$ (red) of the down-up DW and $v^-$ (blue) of the up-down DW, respectively.
Both $v^+$ and $v^-$ show exactly the same inversion symmetry with respect to their symmetry axes, shifted to the opposite directions, shown by the dashed lines.
Such inversion symmetry is attributed to the transition of the DW configurations between two opposite N{\' e}el DWs through Bloch DW.

The opposite shifts of symmetry axes then correspond to the DMI-induced effective magnetic field on DW: $H_{\rm DMI}^+$ of the down-up DW and $H_{\rm DMI}^-$ $(=-H_{\rm DMI}^+)$ of the up-down DW.
The total in-plane field $H_x^*$ on DW can then be defined as $H_x^*=H_x+H_{\rm DMI}^{\pm}$.
Notably, both $v^+$ and $v^-$ are exactly overlapped onto each other, when plotted with respect to $H_x^*$ as shown by the upper panel of Fig.~\ref{fig2}(b).
This observation indicates that the DW configuration is solely determined by $H_x^*$, irrespective of the two DW types.
The exact overlapping is also confirmed for other strips with different $t_{\rm Pt}$, as shown by the upper panels of Figs.~\ref{fig2}(c) and (d).

Due to the same DW configuration for a given $H_x^*$, it is now possible to directly compare the spin-torque efficiencies between the two types of DWs.
The lower panels of Figs.~\ref{fig2}(b)-(d) plot the average $\bar {\varepsilon}(H_x^*)$ (green lines) and the deviation $\delta \varepsilon(H_x^*)$ (purple lines) with respect to $H_x^*$. 
Here, $\bar {\varepsilon}(H_x^*)\equiv[\varepsilon^+(H_x^*)+\varepsilon^-(H_x^*)]/2$ and $\delta \varepsilon(H_x^* )\equiv[\varepsilon^+(H_x^*)-\varepsilon^-(H_x^*)]/2$.
For a reference, ${\varepsilon}^+(H_x^*)$ (red) and ${\varepsilon}^-(H_x^*)$ (blue) are also plotted.
It is interesting that $\bar {\varepsilon}(H_x^*)$ precisely follows $\varepsilon_{\rm SOT}$ (black dashed line) from an analytic prediction, indicating that this contribution is truly attributed to the SOT i.e., $\bar{\varepsilon}=\varepsilon_{\rm SOT}$~\cite{FrankenSciRep, EmoriPRB, comment2}.

However, the most surprising and salient observation is that all the $\delta \varepsilon$ curves exhibit a universal functional shape.
The $\delta \varepsilon$ curves are symmetric with respect to the inversion axis $H_x^*=0$, at which the DW is expected to be in the Bloch wall structure.
This functional shape is totally different from that of the SOT, which is antisymmetric with respect to $H_x^*=0$.
Note that the magnetization canting inside the domains or the DW tilting~\cite{BoullePRL} in the presence of a strong DMI might cause the deviation from the typical $\varepsilon_{\rm SOT}$~\cite{EmoriPRB}.
However, these effects do not meet the observed symmetry of $\delta \varepsilon$ as emphasized in Fig.~\ref{fig2}(b).
Also, these effects are expected to be small, since the present samples show large perpendicular magnetic anisotropy ($\sim8\times10^5$ J/m$^3$)~\cite{KimAPL} and the small DMI-induced field ($\sim25\pm10$ mT).
In addition, the sign of $\delta \varepsilon$ indicates that $\delta \varepsilon$ drives the DWs along the current direction, which is opposite to the prediction of the original STT.
All these observations require another driving force.

\begin{figure}
\includegraphics[width=8.6 cm]{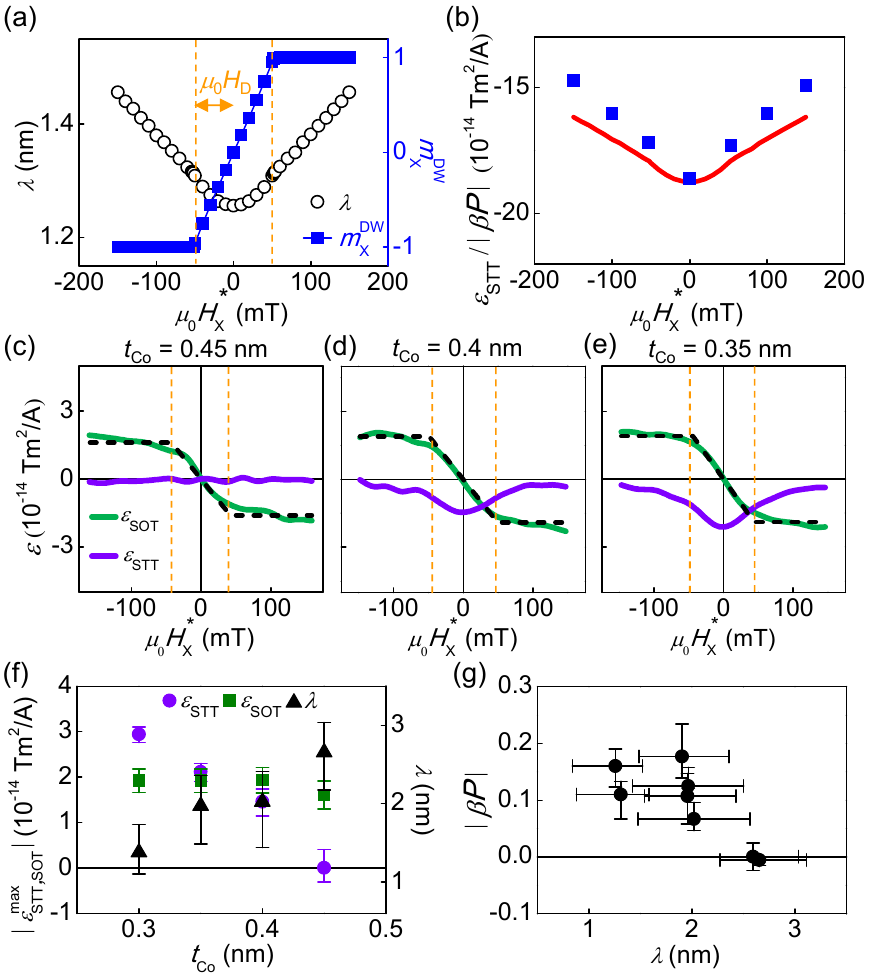}
\caption{
(a) Micromagnetic prediction of $\lambda$ and $m_x^{\rm DW}$ for (2.5-nm Pt/0.3-nm Co/1.5-nm Pt) strip. The orange dashed lines indicate $\mu_0 H_{\rm D}$. (b) Predicted $\varepsilon_{\rm STT}/|\beta P|$ with respect to $\mu_0 H_x^*$. The red line shows the values from the relation $\varepsilon_{\rm STT}/|\beta P|=-\hbar/2eM_{\rm S}\lambda$. The blue symbols show the results from the micromagnetic simulations with the STT module. (c)-(e) Decomposition of $\varepsilon$ into $\varepsilon_{\rm STT}$ and $\varepsilon_{\rm SOT}$ for the strips with $t_{\rm Co}$ of 0.45 (c), 0.4 (d) and 0.35 nm (e). (f) Maximum values of $|\varepsilon_{\rm STT}|$ and $|\varepsilon_{\rm SOT}|$ and estimated $\lambda$ with respect to $t_{\rm Co}$. As a side note, the harmonic Hall measurement yields the consistent results with the $\varepsilon_{\rm SOT}^{\rm max}$ within the same order of magnitude~\cite{comment2}. (g) $|\beta P|$ with respect to $\lambda$. The values were obtained for both the Bloch ($H_x^*=0$) and N{\' e}el ($H_x^*=H_{\rm D}$) configurations in each strip. Because $|P|\le1$, the  $|\beta P|$ presents the lower bound of $|\beta|$.
}
\label{fig3}
\end{figure}

Although, there has been no complete theory predicting $\delta \varepsilon$ observed here, we find that the symmetry and peculiar shape of $\delta \varepsilon(H_x^*)$ are consistent with the behavior caused by the nonadiabatic STT. 
In the simple one-dimensional STT model~\cite{ThiavilleEPLSTT}, the nonadiabatic spin torque efficiency $\varepsilon_{\rm STT}$ is proportional to $\beta P/\lambda$, where $\beta$, $P$, and $\lambda$ denote the nonadiabaticity, spin polarization, and DW width, respectively.
Several mechanisms predict different scalings of the nonadiabaticity.
For example, the ballistic spin-mistracking model~\cite{XiaoPRB} predicts an exponential decay with respect to $\lambda$ and the spin diffusion mechanism~\cite{AkosaPRB} anticipates $1/\lambda^2$ dependence.
Here, we used the $1/\lambda$ dependence for simplicity.
Figure~\ref{fig3}(a) shows the micromagnetic prediction~\cite{comment2, comment3} that $\lambda$ and $m_x^{\rm DW}$ vary with respect to $H_x^*$, where $m_x^{\rm DW}$ is the direction cosine of the magnetization inside the DW along the $x$ axis.
The variation in $\lambda$ and $m_x^{\rm DW}$ are caused by the counterbalance between the magnetostatic energy and the Zeeman energy inside the DW.
The DW demagnetizing field $H_{\rm D}$, which is required to saturate the DW to the N{\' e}el configuration, is depicted in Fig.~\ref{fig3}(a).
Owing to the inverse proportionality of $\varepsilon_{\rm STT}$ on $\lambda$, such $\lambda$ variation results in the peculiar shape of $\varepsilon_{\rm STT}/|\beta P|$ (red line), as shown in Fig.~\ref{fig3}(b).
The $H_x^*$-induced canting of the domains would further enhance the variation in $\varepsilon_{\rm STT}$ as demonstrated by the micromagnetic prediction (blue symbols).

Though the micromagnetic simulation qualitatively reproduces the experimental shape of $\delta \varepsilon(H_x^*)$, the magnitude of the variation is small compared to that of $\delta \varepsilon(H_x^*)$.
Note that this micromagnetic simulation is performed under the condition of a fixed $\beta$.
This may suggest that, in the real situation, the variation in $\beta$ should be considered as a function of $\lambda$.
Nevertheless, it should be emphasized that this simulation results show the symmetry and the shape identical to the experimental $\delta \varepsilon(H_x^*)$ reported here.
This observation, therefore, suggests that the STT can be a possible origin of $\delta \varepsilon$ and hereafter, we will denote $\delta \varepsilon$ as $\varepsilon_{\rm STT}$.

Figures~\ref{fig3}(c)-(e) present the plots of $\varepsilon_{\rm STT}$ (purple) and $\varepsilon_{\rm SOT}$ (green) for (2.5-nm Pt/$t_{\rm Co}$ Co/1.5-nm Pt) strips.
The shape and the magnitude of $\varepsilon_{\rm SOT}$ in these strips do not exhibit notable change with $t_{\rm Co}$, indicating that the SOT is well controlled in these strips~\cite{comment6}.
However, the maximum of $|\varepsilon_{\rm STT}|$ radically increases and, eventually exceeds the maximum of $|\varepsilon_{\rm SOT}|$ as $t_{\rm Co}$ decreases down to 0.35 nm.
As shown in Fig.~\ref{fig2}, all of the strips with $t_{\rm Co}=0.3$ nm also exhibit the maximum of $|\varepsilon_{\rm STT}|$ is larger than the maximum of $|\varepsilon_{\rm SOT}|$.
One can therefore conclude that $\varepsilon_{\rm STT}$ rather than $\varepsilon_{\rm SOT}$ provides the major driving force responsible for the CIDWM in these strips.
Figure~\ref{fig3}(f) summarizes the measured maximum values of both contributions, $\varepsilon_{\rm STT}^{\rm max}$ and $\varepsilon_{\rm SOT}^{\rm max}$, with respect to $t_{\rm Co}$.
With the large $\varepsilon_{\rm STT}^{\rm max}$, the STT generates the effective field up to $\sim3$ mT for $J$=$10^{11}$ A/m$^2$.
We note that this value is even comparable to that of SOT ($\sim4$ mT for the same $J$) in Pt/Co/Oxide trilayers with uncompensated spin Hall currents~\cite{FrankenSciRep, PaiPRB}.

In the framework of the nonadiabatic STT, such significant enhancement of $\varepsilon_{\rm STT}$ might be the consequence of the ultra-narrow $\lambda$~\cite{TataraPRL}.
It is worth noting that our films exhibit large perpendicular magnetic anisotropy $K_{\rm U}$~\cite{comment2, KimAPL}, whereas the exchange stiffness $A_{\rm x}$  is largely reduced as indicated by the reduction of the Curie temperature~\cite{comment2}.
Such reduction of $A_{\rm x}$  was generally observed in a-few-monolayers-thick films~\cite{PratzerPRL, EyrichPRB, NembachNatPhys, MoreauNatNano}.
It is therefore natural to expect a narrow DW width, which is known to be proportional to $\sqrt{A_{\rm x}/K_{\rm U}}$. 
Moreover, the obsevation of large $H_{\rm D}$ [orange dashed lines in Fig.~\ref{fig2}(c) and Figs.~\ref{fig3}(c)-(e)] supports the formation of the ultra-narrow DWs in our films.
Since  $H_{\rm D}$ is proportional to the DW demagnetizing factor between the Bloch and N{\' e}el DWs, 3$\sim$5-times larger $H_{\rm D}$ and $\sim$2-times thinner $t_{\rm Co}$ corresponds to 6$\sim$10-times narrower $\lambda$ in comparison to the other materials~\cite{FrankenSciRep, EmoriPRB}.
The black triangles in Fig.~\ref{fig3}(f) show the estimated $\lambda$ from the $H_{\rm D}$ measurement~\cite{comment5, TarasenkoJMMM}.
The results show that $\lambda$ becomes ultra-narrow down to a few nanometers as $t_{\rm Co}$ decreases.
Note that thermal effects that might impact the DW width are not taken into consideration for this estimate.
Furthermore, as it has been shown that the STT efficiency can exhibit a temperature dependence~\cite{LaufenbergPRL, BoullePRL, DevolderJAP}, our results derived from a simple temperature-independent model description can only provide qualitative information on the overall trend of $\lambda$.
Additional micromagnetic simulation about the influence of thermal fluctuation on $\lambda$ can be found in Ref.~\cite{comment2}.

For the case $\lambda$ is comparable to the transport scale of about a few nanometers such as the spin-diffusion and Larmor precession lengths, it has been theoretically predicted that $\beta$ should exhibit large variation, depending on $\lambda$~\cite{BohlensPRL, VanhaverbekePRB}.
The variation in $\beta$ was conjectured by use of the relation $\beta P=(2eM_{\rm S}/ℏ\hbar)\varepsilon_{\rm STT}\lambda$ with the Planck constant $\hbar$, electron charge $e$, and saturation magnetization $M_{\rm S}$~\cite{ThiavilleEPLSTT} as shown in Fig.~\ref{fig3}(g).
The plot shows drastic variation of $\beta P$ in a narrow range of $\lambda$.
Although this analysis is based on simple one-dimensional STT model, this observation signals the possibility of engineering $\varepsilon_{\rm STT}$ for further enhancement by tuning the magnetic properties such as the interfacial anisotropy and the exchange stiffness.

\begin{figure}
\includegraphics[width=8.6 cm]{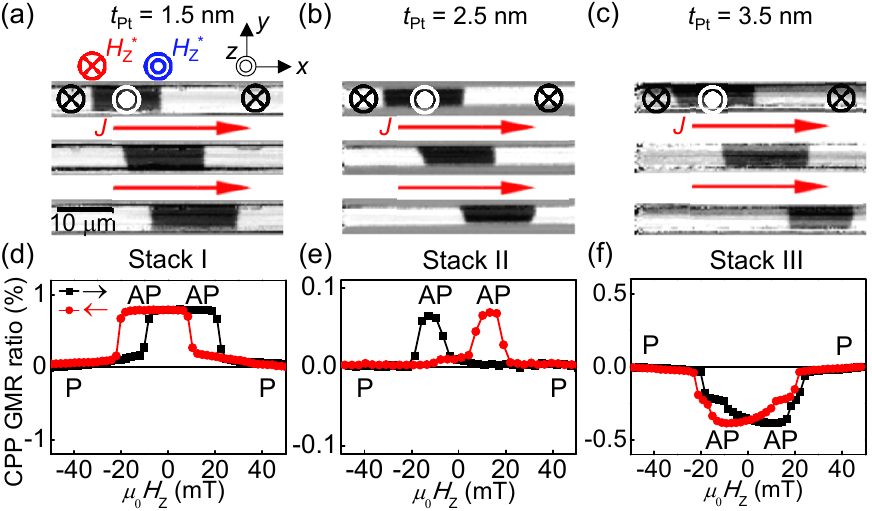}
\caption{
(a)-(c) Scanned Kerr images of the purely current-induced DW displacements in the (2.5-nm Pt/0.3-nm Co/$t_{\rm Pt}$ Pt) strips with $t_{\rm Pt}= 1.5$ (a), 2.5 (b) and 3.5 nm (c). The red (blue) symbol indicates the direction of $H_z^*$ inside the down-up (up-down) DW. (d)-(f) CPP-GMR curves of Stacks I (d), II (e) and III (f). The parallel and anti-parallel states are indicated by P and AP, respectively.
}
\label{fig4}
\end{figure}

Finally, the sign of $\varepsilon_{\rm STT}$ is examined.
As previously mentioned, such negative sign of $\varepsilon_{\rm STT}$ induces the DW motion along the current flow.
This is experimentally confirmed that all the DWs move along the current direction irrespective of the sign of the SOT as shown in Figures~\ref{fig4}(a)-(c)~\cite{comment4}.
The negative sign of $\varepsilon_{\rm STT}$ can be a consequence of either a negative $P$ or a negative $\beta$ as previously discussed in several reports~\cite{LeePRL, MironNatMater, comment7, GaratePRB, GilmorePRB}.
In contrast with the presumption of positive parameters in the STT theory, theoretical studies have revealed the possibilities of a negative $P$ in CoPt alloys with dilute Co concentration~\cite{SiprPRB} and a negative $\beta$ in DWs narrower than several nanometers~\cite{BohlensPRL}.
Our experimental situation might be relevant to the cases of a dilute Co concentration with very thin Co layer sandwiched by thicker Pt layers and/or the ultra-narrow DWs. 

To examine these scenarios, we performed a giant magnetoresistance (GMR) measurement in the geometry of the current perpendicular to the plane (CPP)~\cite{HsuPRL}. 
Three distinct spin valve stacks with 5.0-nm Ta/2.5-nm Pt/($t_{\rm A}$ Co/3.0-nm Pt/$t_{\rm B}$ Co/3.0-nm Pt)$_2$ were patterned into dots.
The Stack I consists of both thick Co layers ($t_{\rm A}= 1.4$ nm and $t_{\rm B}= 1.5$ nm) and the Stack II consists of both thin Co layers ($t_{\rm A}=0.4$ nm and $t_{\rm B}=0.3$ nm), whereas the Stack III is a combination of thick and thin Co layers ($t_{\rm A}=0.3$ nm and $t_{\rm B}= 1.4$ nm).
Figures~\ref{fig4}(d)-(f) summarize the measurement results.
The Stacks I and II exhibit positive CPP-GMRs with higher resistance at the antiparallel state. However, interestingly, the Stack III exhibits an inverse CPP-GMR.
Because all other layer structures were kept the same except the Co layer thicknesses, the inverse CPP-GMR might originate from the different transport properties of the Co layers.
This result suggests that a negative $P$ appears in the strips with $t_{\rm Co}\le0.4$ nm, indicating that a negative $P$, rather than a negative $\beta$, is responsible for the negative $\varepsilon_{\rm STT}$~\cite{comment2}.
Although these observations show remarkably new features in the atomically thin Co layer, however, further studies, such as the origin of the negative spin polarization, crystallography of the atomically thin Co layer sandwiched by Pt layers and the additional effect from adjacent Pt layers~\cite{RyuNatComm}, are required.

In contrast to the present consensus that the STT vanishes in a thin ferromagnetic layer, here, we showed that the nonadiabatic STT is significantly enhanced so that it is even larger than the SOT in ultra-thin Pt/Co/Pt strips, and consequently, the CIDWM is governed by the STT.
Such significantly enhanced STT is caused by the formation of the ultra-narrow DW of a few nanometers, which induces a radical increase of the nonadiabaticity.
Moreover, it is found that the STT is negative, resulting in the DW motion along the current flow.
All these observations imply the controllability of the nonadiabatic STT efficiency and thus, promise the emerging DW-mediated logic and memory devices.

This work was supported by the National Research Foundations of Korea (NRF) grant funded by the Ministry of Science, ICT and Future Planning of Korea (MSIP) (2015R1A2A1A05001698, 2015M3D1A1070476). 
S.-G.J. was supported by a NRF grant funded by the Korea government (NRF-2011-Global Ph.D. Fellowship Program).
B.-C.M. was supported by the KIST institutional program and Pioneer Research Center Program of MSIP/NRF (2011-0027905).

\newpage

\newpage
\newpage
\newpage


\begin{thebibliography}{99}
\bibitem{comment1} Cancellation of SOT in this symmetric structure is also confirmed by the in-plane current-induced magnetization switching experiment. In this strip, the switching is not observed for even more longer and higher current pulses. For more details, see~\cite{comment2}.
\bibitem{comment5} DW width $\lambda$ is estimated by the relation $\lambda\approx2ln2\cdot t_{\rm f}M_{\rm S}/\pi^2H_{\rm D}$~\cite{TarasenkoJMMM}, where the present $\lambda$ corresponds to the DW width at the N{\' e}el configuration, $M_{\rm S}$ is saturation magnetization and $t_{\rm f}$ is magnetic layer thickness. A micromagnetic calculation confirms the relation within an accuracy of 10 ${\%}$~\cite{comment2}.
\bibitem{comment4} Note that the signs of SOT are opposite for $t_{\rm Pt}=1.5$ nm and $t_{\rm Pt}=3.5$ nm~\cite{comment2}. Since the favored chirality in both samples are found to be the same (from the asymmetric DW speed measurement), SOT mechanism should result in opposite DW motions. Moreover, for $t_{\rm Pt}=2.5$ nm, due to the vanishingly small SOT, one would expect no (or slow) motion in the framework of SOT mechanism. However, the predictions are completely inconsistent with the experimental observations. These observations support again the presence of negative nonadiabatic STT component.
\bibitem{comment7} The negative nonadiabaticity has been also predicted by the mechanisms such as spin-orbit coupling and spin-dependent scattering~\cite{GaratePRB, GilmorePRB}, but they give rise to contributions independent of $\lambda$.
\end{thebibliography}
\end{document}